\documentclass{jpsj2}

\begin{document}
\title{
Generalized Phase Rules
}

\author{Akira Shimizu \thanks{E-mail address: shmz@ASone.c.u-tokyo.ac.jp}}
\inst{
Department of Basic Science, University of Tokyo, 
3-8-1 Komaba, Tokyo 153-8902, Japan
}
\date{\today}
\abst{
For a multi-component system, 
general formulas are derived for the dimension
of a coexisting region in the phase diagram 
in various state spaces.
}
\kword{
phase diagram, phase boundary, coexisting region, thermodynamics, multi components, phase transition, phase rule, new materials
}

\maketitle

\section{Introduction}
\label{sec:intro}

The first-order phase transition 
plays an important role in diverse fields of physics.
Its most characteristic feature is that 
a phase boundary 
is a coexisting region,
where two or more phases coexist.
%
%
The dimension $D$ of a coexisting region in 
the phase diagram depends on the number $r$ of coexisting 
phases \cite{Gibbs,Landau,AS,Callen,Fermi}.
It also depends on {\em which variables are taken} 
as the axes of the phase diagram \cite{Gibbs,Landau,AS}.

\begin{figure}[htb]
\begin{center}
\includegraphics[width=0.8\linewidth]{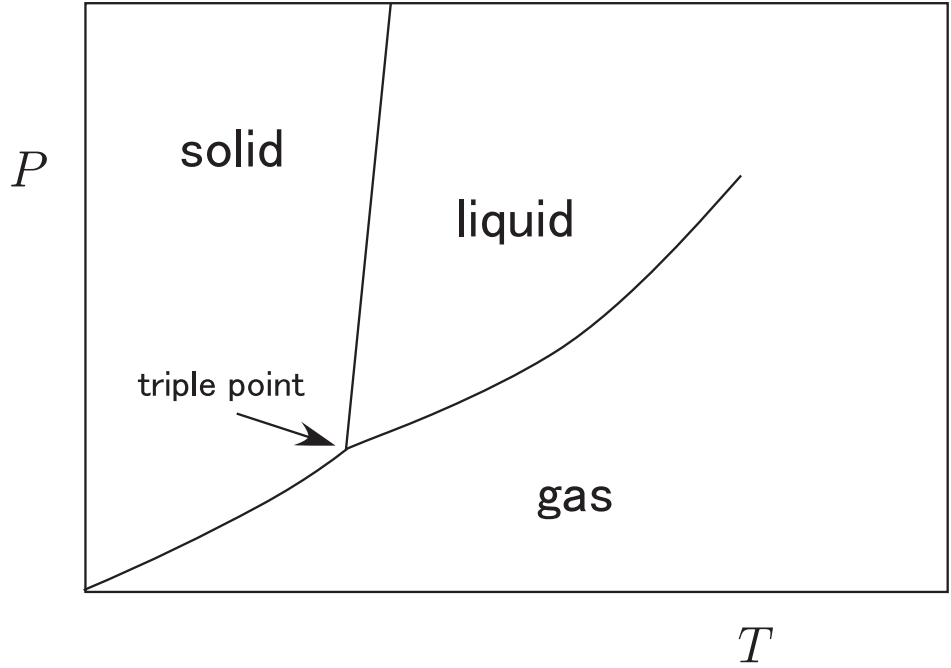}
\end{center}
\caption{
A schematic phase diagram of a single-component system, 
plotted in the space spanned by the axes corresponding to $T$ and $P$.
The solid-liquid-gas coexisting region
is a point, which is called the triple point.
On the other hand,
the liquid-gas, solid-liquid and solid-gas coexisting regions are 
lines, which are called  coexisting lines.
}
\label{fig:PD}
\end{figure}

\begin{figure}[htb]
\begin{center}
\includegraphics[width=0.8\linewidth]{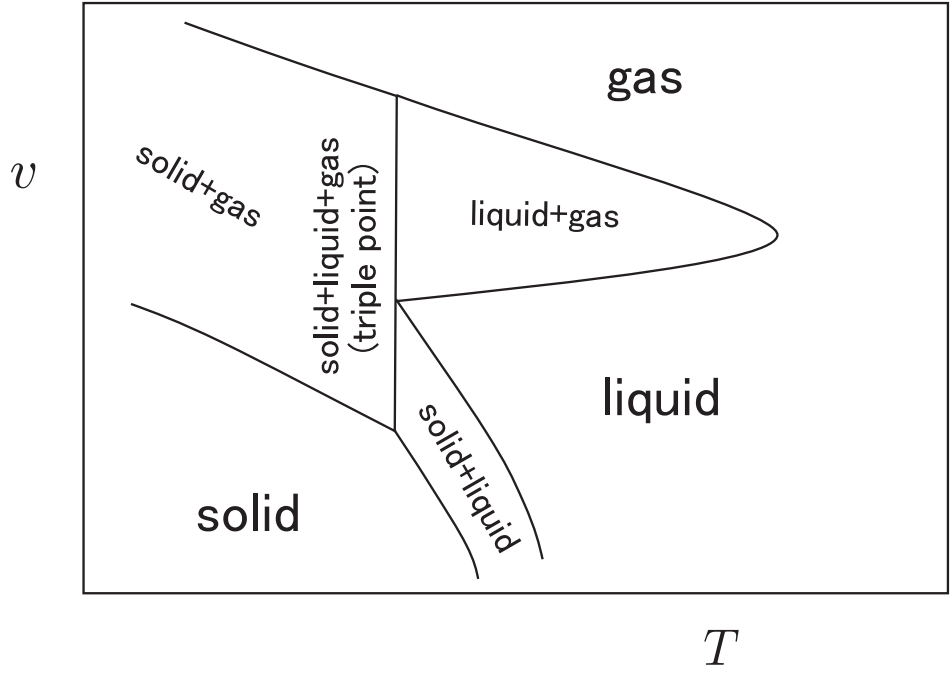}
\end{center}
\caption{
A schematic phase diagram of a single-component system, 
plotted in the space spanned by the axes corresponding to $T$ and $v$ $(\equiv V/N)$.
Unlike Fig.~\ref{fig:PD}, 
the solid-liquid-gas coexisting region (triple point)
is a vertical line, whereas
the liquid-gas, solid-liquid and solid-gas coexisting regions are 
planes.
}
\label{fig:PD_TV}
\end{figure}

\begin{figure}[htb]
\begin{center}
\includegraphics[width=0.8\linewidth]{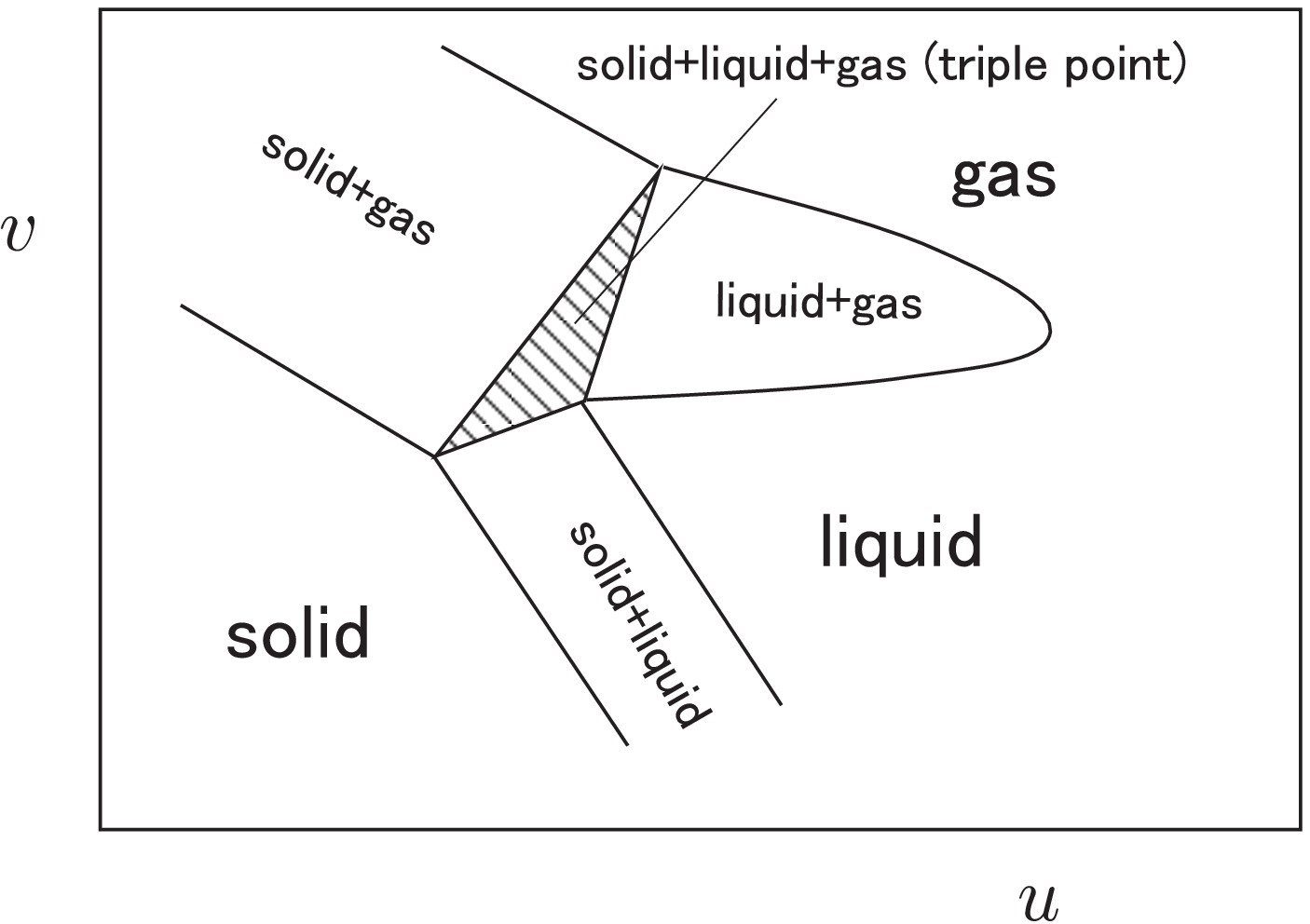}
\end{center}
\caption{
A schematic phase diagram of a single-component system, 
plotted in the space spanned by the axes corresponding to $u$ $(\equiv U/N)$ and $v$ $(\equiv V/N)$.
All the coexisting regions are planes.
For example, the triple point,
which was a point in Fig.~\ref{fig:PD}, 
extends to the shaded area in this figure.
Unlike Figs.~\ref{fig:PD} and \ref{fig:PD_TV}, 
every point in this diagram
corresponds to a single equilibrium state,
and different points correspond to different equilibrium states.
}
\label{fig:PD_UV}
\end{figure}

For example,
for a single-component system whose
natural variables of the entropy \cite{Gibbs,Landau,AS,Callen}
are $U$ (energy), $V$ (volume) and $N$ (amount of substance), 
the phase diagram may be drawn either in the $T$-$P$ plane,
or in the $T$-$v$ plane, 
or in the $u$-$v$ plane, 
where $T, P, v$ and $u$ are the 
temperature, pressure, 
molar volume ($\equiv V/N$) and molar energy ($\equiv U/N$), 
respectively.
Here, $v$ and/or $u$ are used instead of $V$ and/or $U$
in order to make the diagrams independent of $N$.
These phase diagrams are schematically shown in 
Figs.~\ref{fig:PD}, \ref{fig:PD_TV} and \ref{fig:PD_UV}, 
respectively \cite{AS}.
We denote $D$ for these diagrams by $D(T,P)$, $D(T,v)$ and $D(u,v)$, 
respectively.
For example, 
$D(T,P)=1, D(T,v)=D(u,v)=2$ for a liquid-gas coexisting region, 
where two phases coexist (i.e., $r=2$), 
whereas 
$D(T,P)=0, D(T,v)=1, D(u,v)=2$ 
at the triple point, where $r=3$.\cite{Gibbs,Landau,AS}
Since $T$ and $P$ can take constant values
across a coexisting region,\cite{Gibbs,Landau,AS}
$D$ of the coexisting region tends to shrink
if $T$ and/or $P$ is taken as an axis(es) of the phase diagram,
and thus $D(T,P) \leq D(T,v) \leq D(u,v)$.

For such a single-component system, 
$D(T,P)$ coincides with the 
`thermodynamical degrees of freedom' $f$, 
which is given by 
the Gibbs phase rule \cite{Gibbs,Landau,AS,Callen,Fermi}
(see eq.~(\ref{eq:Gibbs}) below).
On the other hand, $D(u,v) \geq D(T,v) \geq f$
in general \cite{Gibbs,AS}.
For example, 
$ 
f=1=D(T,P) < D(T,v)=D(u,v)=2
$ 
for a liquid-gas coexisting region, 
whereas 
$ 
f=0=D(T,P)< D(T,v)< D(u,v)=2
$ 
at the triple point.

The situation becomes more complicated for a multi-component system,
which consists of $q$ ($\geq 2$) different substances.
We assume that 
the natural variables of entropy are $U,V,N_1, \ldots, N_q$,
where $N_j$ ($j=1, \ldots, q$) is the amount of substance $j$.
That is, the entropy $S$ is a function of these 
additive (extensive) variables;
\begin{equation}
S = S(U,V,N_1, \ldots, N_q),
\label{eq:FR}\end{equation}
which is called the fundamental relation \cite{Gibbs,AS,Callen}.
Let $N_{\rm tot}$ be the total amount of substances, 
\begin{equation}
N_{\rm tot} \equiv N_1 + \ldots+ N_q.
\end{equation}
To make phase diagrams independent of $N_{\rm tot}$, 
we use the normalized variables;
\begin{eqnarray}
u &\equiv& {U \over N_{\rm tot}},
\\
v &\equiv& {V \over N_{\rm tot}},
\\
n_j &\equiv& {N_j \over N_{\rm tot}} \quad (j=1, \ldots, q).
\end{eqnarray}
Among $n_1, \ldots, n_q$, only $(q-1)$ variables are independent because
of the identity
\begin{equation}
\sum_{j=1}^q n_j=1.
\label{eq:sum_nj}\end{equation}
Hence, the phase diagram may be drawn 
in the $(q+1)$-dimensional space that 
is spanned by the axes corresponding to either $(T, P, n_1, \ldots, n_{q-1})$, 
$(T, v, n_1, \ldots, n_{q-1})$,
or $(u, v, n_1, \ldots, n_{q-1})$,
which correspond to the so-called 
$TPN$, $TVN$ and $UVN$ representations, respectively.
We denote the dimensions of a coexisting region 
in these spaces by 
$D(T, P, n_1, \ldots, n_{q-1})$, 
$D(T, v, n_1, \ldots, n_{q-1})$
and $D(u, v, n_1, \ldots, n_{q-1})$, respectively. 

On the other hand, 
the thermodynamical degrees of freedom\cite{Gibbs,Landau,AS,Callen,Fermi}
$f$ is defined as the number of variables
that can be varied independently in the coexisting region,
among the intensive variables $T, P, \mu_1, \ldots, \mu_q$, 
where $\mu_j$ denotes the chemical potential of
substance $j$.
The Gibbs phase rule gives \cite{Gibbs,Landau,AS,Callen,Fermi}
\begin{equation}
f=q+2-r.
\label{eq:Gibbs}\end{equation}
For $q \geq 2$ neither of $D$'s coincides with $f$ in general,
and the explicit formulas for $D$'s were unknown.

Knowledge about $D$'s would be very helpful in 
drawing phase diagrams of new materials \cite{Gibbs,Landau,AS,Callen,Fermi}.
Since phase diagrams are most fundamental to studies of macroscopic 
systems, 
general formulas would be valuable which give
$D(T, P, n_1, \ldots, n_{q-1})$, 
$D(T, v, n_1, \ldots, n_{q-1})$ and 
$D(u, v, n_1, \ldots, n_{q-1})$ 
as functions of $q$ and $r$.
The purpose of the present paper is to derive such 
general formulas.
We also evaluate $D$'s in some other spaces.
The results will be summarized as the chain of equalities and inequalities, 
eq.~(\ref{eq:chain}).

\section{$UVN$ representation}

\subsection{Basic relations}
We consider a $q$-component system, 
the natural variables of whose entropy \cite{Gibbs,Landau,AS,Callen} 
are assumed to be $U,V,N_1, \ldots, N_q$.
For each coexisting region, 
we label the coexisting phases by $\alpha=1, \ldots, r$.
For example, the shaded region of Fig.~\ref{fig:PD_UV}
is a coexisting region, in which 
the gas ($\alpha=1$), liquid  ($\alpha=2$) and solid ($\alpha=3$)
phases coexist, hence is called the triple point.
Note that there is the trivial $r!$-fold degeneracy 
in labeling $\alpha=1,\ldots,r$.
Since this degeneracy does not affect the 
values of $D$'s,
we henceforth forget about it.

The values of $U, V$ and $N_j$ 
in phase $\alpha$ are denoted by
$U^\alpha,V^\alpha$ and $N^\alpha_j$, respectively.
They are related to those of the total system as
\begin{eqnarray}
U &=& \sum_{\alpha=1}^r U^\alpha,
\label{eq:U=sumUk}\\
V &=& \sum_{\alpha=1}^r V^\alpha,
\label{eq:V=sumVk}\\
N_j &=& \sum_{\alpha=1}^r N^\alpha_j \quad (j=1, \ldots, q).
\label{eq:N=sumNk}
\end{eqnarray}
Corresponding to $u, v, n_j$ of the total system,
we define 
\begin{eqnarray}
u^\alpha &\equiv& {U^\alpha \over N^\alpha_{\rm tot}},
\\
v^\alpha &\equiv& {V^\alpha \over N^\alpha_{\rm tot}},
\\
n^\alpha_j &\equiv& {N^\alpha_j \over N^\alpha_{\rm tot}} 
\quad (j=1, \ldots, q)
\end{eqnarray}
for each phase, 
where $N^\alpha_{\rm tot}$ is the total amount 
of substances in phase $\alpha$, 
\begin{equation}
N^\alpha_{\rm tot} \equiv N^\alpha_1 + \ldots+ N^\alpha_q
\quad (\alpha=1, \ldots, r).
\end{equation}
Corresponding to eq.~(\ref{eq:sum_nj}), we have
\begin{equation}
\sum_{j=1}^q n^\alpha_j =1
\ \mbox{ for all $\alpha$}.
\end{equation}
We also define the molar fraction of each phase
\begin{equation}
\nu^\alpha \equiv {N^\alpha_{\rm tot} \over N_{\rm tot}} 
\quad (\alpha=1, \ldots, r)
\end{equation}
which satisfies the trivial identity
\begin{equation}
\sum_{\alpha=1}^r \nu^\alpha =1.
\label{eq:sum-tnk}\end{equation}

By dividing eqs.~(\ref{eq:U=sumUk})-(\ref{eq:N=sumNk}) by $N_{\rm tot}$,
we obtain
\begin{eqnarray}
u &=& \sum_{\alpha=1}^r \nu^\alpha u^\alpha,
\label{eq:u=sumtnkuk}\\
v &=& \sum_{\alpha=1}^r \nu^\alpha v^\alpha,
\label{eq:v=sumtnkvk}\\
n_j &=& \sum_{\alpha=1}^r \nu^\alpha n_j^\alpha
\quad (j=1, \ldots, q).
\label{eq:n=sumtnknk}\end{eqnarray}
Note that {\em each} phase is, by definition \cite{AS,Fermi,def:phase}, 
homogeneous spatially.
Hence, 
the {\em local} value of the molar energy is equal to 
$u^\alpha$ at everywhere in phase $\alpha$.
On the other hand, the {\em total} system is inhomogeneous 
in a coexisting region,
and the {\em local} value of the molar energy is equal not to $u$ but 
to either of $u^1, u^2, \ldots$, or $u^r$.
As seen from eq.~(\ref{eq:u=sumtnkuk}), 
$u$ actually represents the {\em average} molar energy, which 
is the weighted average of $u^\alpha$'s of the coexisting phases.
The same can also be said about $v$ and $n_j$.

Intensive variables in phase $\alpha$ are continuous functions of 
$u^\alpha,v^\alpha,n^\alpha_1, \ldots, n^\alpha_{q-1}$ \cite{Gibbs,AS},
and have the same values as those of the total system \cite{Gibbs,Landau,AS,Callen};
\begin{eqnarray}
&& T=T(u^\alpha,v^\alpha,n^\alpha_1, \ldots, n^\alpha_{q-1})
=T(u,v,n_1, \ldots, n_{q-1}),\\
&& P=P(u^\alpha,v^\alpha,n^\alpha_1, \ldots, n^\alpha_{q-1})
=P(u,v,n_1, \ldots, n_{q-1}),\\
&& \mu_j=\mu_j(u^\alpha,v^\alpha,n^\alpha_1, \ldots, n^\alpha_{q-1})
=\mu_j(u,v,n_1, \ldots, n_{q-1})
 \quad (j=1, \ldots, q).
\end{eqnarray}
Here, the functions $T(\cdot, \cdot, \ldots, \cdot)$, 
$P(\cdot, \cdot, \ldots, \cdot)$
and $\mu_j(\cdot, \cdot, \ldots, \cdot)$
are obtained by differentiating the fundamental relation, 
eq.~(\ref{eq:FR}) \cite{Gibbs,AS,Callen}.

\subsection{$D$ in the 
$\{ u^\alpha,v^\alpha,n^\alpha_1, \ldots, n^\alpha_{q-1}, \nu^\alpha \}_{\alpha=1,\ldots,r}$
space}

Our principal purpose is to evaluate 
$D$'s in the $(q+1)$-dimensional spaces such as
the one spanned by the axes corresponding to $(T, P, n_1, \ldots, n_{q-1})$.
As a preliminary, 
we first evaluate the dimension 
$D(\{ u^\alpha,v^\alpha,n^\alpha_1, \ldots, n^\alpha_{q-1}, \nu^\alpha \}_\alpha)$ 
in a larger space which is spanned by the axes corresponding to 
$r(q+2)$ variables
$\{ u^\alpha,v^\alpha,n^\alpha_1, \ldots, 
n^\alpha_{q-1}, \nu^\alpha_{\rm tot} \}_{\alpha=1,\ldots,r}$ \cite{dependent}.

Since $r$ different phases coexist, we have 
\begin{eqnarray}
&&
T(u^1,v^1,n^1_1, \ldots, n^1_{q-1})=\ldots=
T(u^r,v^r,n^r_1, \ldots, n^r_{q-1}),
\\
&&
P(u^1,v^1,n^1_1, \ldots, n^1_{q-1})=\ldots=
P(u^r,v^r,n^r_1, \ldots, n^r_{q-1}),
\\
&&
\mu_j(u^1,v^1,n^1_1, \ldots, n^1_{q-1})=\ldots=
\mu_j(u^r,v^r,n^r_1, \ldots, n^r_{q-1}) \quad (j=1, \ldots, q),
\end{eqnarray}
which impose $(r-1) \times (q+2)$ conditions
on $r \times (q+1)$ variables 
$\{ u^\alpha,v^\alpha,n^\alpha_1, \ldots, n^\alpha_{q-1} \}_{\alpha=1,\ldots,r}$.
Therefore, 
the dimension of the set of values of 
$\{ u^\alpha,v^\alpha,n^\alpha_1, \ldots, n^\alpha_{q-1} \}_{\alpha=1,\ldots,r}$
is evaluated as
\begin{equation}
r(q+1)-(r-1)(q+2)=q+2-r.
\end{equation}
Among the residual variables 
$\nu^1, \ldots, \nu^r$,
we have eq.~(\ref{eq:sum-tnk}).
Hence, 
\begin{eqnarray}
D(\{ u^\alpha,v^\alpha,n^\alpha_1, \ldots, n^\alpha_{q-1}, 
\nu^\alpha \}_\alpha)
&=&
(q+2-r)+(r-1)
\nonumber\\
&=&
q+1.
\label{eq:Duvn_a}\end{eqnarray}

\subsection{$D$ in the $u,v,n_1, \ldots, n_{q-1}$ space}

With the help of eq.~(\ref{eq:Duvn_a}), 
we can evaluate $D(u,v,n_1, \ldots, n_{q-1})$
as follows.
It is obvious from eqs.~(\ref{eq:U=sumUk})-(\ref{eq:N=sumNk}) 
that the values of 
$U,V,N_1, \ldots, N_q$ are uniquely determined by 
the values of $\{ U^\alpha,V^\alpha,N_1^\alpha, \ldots, N_q^\alpha \}_{\alpha=1,\ldots,r}$.
Furthermore, 
the latter (including the number $r$) 
is uniquely determined by the former, 
because otherwise two different states would have the same value of the 
total entropy $S(U,V,N_1, \ldots, N_q)$, and thus the total system 
would be unstable \cite{AS}.
Therefore, 
the values of $\{ U^\alpha,V^\alpha,N_1^\alpha, \ldots, N_q^\alpha \}_{\alpha=1,\ldots,r}$ (including $r$) 
have one-to-one correspondence with 
the values of $U,V,N_1, \ldots, N_q$ of the total system.
This means that 
$\{ u^\alpha,v^\alpha,n^\alpha_1, \ldots, n^\alpha_{q-1}, N^\alpha_{\rm tot} \}_{\alpha=1,\ldots,r}$ 
have one-to-one correspondence with 
$u,v,n_1, \ldots, n_{q-1}, N_{\rm tot}$.
Hence, 
$\{ u^\alpha,v^\alpha,n^\alpha_1, \ldots, n^\alpha_{q-1}, \nu^\alpha \}_{\alpha=1,\ldots,r}$
have one-to-one correspondence with 
$u,v,n_1, \ldots, n_{q-1}$.
Therefore, 
\begin{eqnarray}
D(u,v,n_1, \ldots, n_{q-1})
&=&
D(\{ u^\alpha,v^\alpha,n^\alpha_1, \ldots, n^\alpha_{q-1}, \nu^\alpha \}_\alpha)\nonumber\\
&=&
q+1,
\label{eq:Duvn}\end{eqnarray}
which is equal to the dimension of 
all possible states for a given value of $N_{\rm tot}$ \cite{Gibbs,AS,Callen}.
This is reasonable (obvious in some sense) because
each equilibrium state
has one-to-one correspondence with the set of values of 
$U,V,N_1, \ldots, N_q$,
even when the equilibrium state is in a coexisting region \cite{Gibbs,AS}.

For a single-component system, for example, formula (\ref{eq:Duvn}) gives
$ 
D(u,v)=2,
$ 
which is consistent with Fig.~\ref{fig:PD_UV}.

\section{$TVN$ representation}

\subsection{$D$ in the 
$T,\{ v^\alpha, n^\alpha_1, \ldots, n^\alpha_{q-1} \}_{\alpha=1,\ldots,r}$
space}

To evaluate $D(T,v,n_1, \ldots, n_{q-1})$, 
we start with considering $D$ in a larger space which is spanned by
the axes corresponding to 
$(T,\{ v^\alpha, n^\alpha_1, \ldots, n^\alpha_{q-1} \}_{\alpha=1,\ldots,r})$
\cite{dependent}.

Since $r$ different phases coexist, we have 
\begin{eqnarray}
&&
P^1(T,v^1,n^1_1, \ldots, n^1_{q-1})=\ldots=
P^r(T,v^r,n^r_1, \ldots, n^r_{q-1}),
\\
&&
\mu_j^1(T,v^1,n^1_1, \ldots, n^1_{q-1})=\ldots=
\mu_j^r(T,v^r,n^r_1, \ldots, n^r_{q-1}) \quad (j=1, \ldots, q),
\end{eqnarray}
which impose $(r-1) \times (q+1)$ conditions
on $(1+r \times q)$ variables.
Therefore, 
\begin{eqnarray}
D(T,\{ v^\alpha, n^\alpha_1, \ldots, n^\alpha_{q-1} \}_\alpha)
&=&
1+rq-(r-1)(q+1)
\nonumber\\
&=&
q+2-r,
\label{eq:DTvn_a}\end{eqnarray}
which coincides with $f$.

\subsection{$D$ in the $T,v,n_1, \ldots, n_{q-1}$ space}

With the help of eq.~(\ref{eq:DTvn_a}), 
we can evaluate $D(T,v,n_1, \ldots, n_{q-1})$ as follows.
Equations (\ref{eq:v=sumtnkvk}) and (\ref{eq:n=sumtnknk}) show that
for each set of values of
$(T,\{ v^\alpha, n^\alpha_1, \ldots, n^\alpha_{q-1} \}_{\alpha=1,\ldots,r})$
one can vary $q$ variables
$v, n_1, \ldots, n_{q-1}$ by varying 
$r$ variables $\{ \nu^\alpha \}_{\alpha=1,\ldots,r}$ 
subject to one condition, eq.~(\ref{eq:sum-tnk}).
Therefore,
\begin{eqnarray}
D(T,v,n_1, \ldots, n_{q-1})
&=&
D(T,\{ v^\alpha, n^\alpha_1, \ldots, n^\alpha_{q-1} \}_\alpha)
+\min \{ q,r-1 \}
\nonumber\\
&=&
q+2-r +\min \{q,r-1 \}
\\
&=&
\begin{cases}
q+1 & (r \leq q+1), \\
q & (r = q+2),
\end{cases}
\label{eq:DTvn}
\end{eqnarray}
where in the last line we have taken account of 
an important consequence of eq.~(\ref{eq:Gibbs}); 
\begin{equation}
r \leq q+2,
\label{eqw:rleqq+2}\end{equation}
which gives the upper limit of the number of coexisting 
phases  \cite{Gibbs,Landau,AS,Callen}.

For a single-component system ($q=1$), for example, the above formula gives
\begin{equation}
D(T,v)
=
\begin{cases}
2 & (r \leq 2), \\
1 & (r =3),
\end{cases}
\end{equation}
which is consistent with Fig.~\ref{fig:PD_TV}.

\section{$TPN$ representation}

\subsection{$D$ in the 
$T,P,\{ n^\alpha_1, \ldots, n^\alpha_{q-1} \}_{\alpha=1,\ldots,r}$ 
space}

To evaluate $D(T,P,n_1, \ldots, n_{q-1})$,
we first consider $D$ in a larger space which is spanned by
the axes corresponding to 
$(T,P,\{ n^\alpha_1, \ldots, n^\alpha_{q-1} \}_{\alpha=1,\ldots,r})$
\cite{dependent}.

Since $r$ different phases coexist, we have 
\begin{equation}
\mu_j^1(T,P,n^1_1, \ldots, n^1_{q-1})=\ldots=
\mu_j^r(T,P,n^r_1, \ldots, n^r_{q-1}) \quad (j=1, \ldots, q)
\label{eq:muTP}\end{equation}
which impose $(r-1) \times q$ conditions
on $[ 2+r \times (q-1) ]$ variables.
Therefore, 
\begin{eqnarray}
D(T,P,\{ n^\alpha_1, \ldots, n^\alpha_{q-1} \}_\alpha)
&=&
2+r(q-1)-(r-1)q
\nonumber\\
&=&
q+2-r,
\label{eq:Gibbs2}\end{eqnarray}
which coincides with $f$. 
This well-known result is also called the 
phase rule \cite{AS,Callen}.

\subsection{$D$ in the $T,P,n_1, \ldots, n_{q-1}$ space}

It is seen from (\ref{eq:n=sumtnknk}) that
for each set of values of
$(T,P,\{ n^\alpha_1, \ldots, n^\alpha_{q-1} \}_{\alpha=1,\ldots,r})$,
one can vary $(q-1)$ variables
$n_1, \ldots, n_{q-1}$ by varying 
$r$ variables $\{ \nu^\alpha \}_{\alpha=1,\ldots,r}$ 
subject to one condition, eq.~(\ref{eq:sum-tnk}).
Therefore,
\begin{eqnarray}
D(T,P,n_1, \ldots, n_{q-1})
&=&
D(T,P,\{ n^\alpha_1, \ldots, n^\alpha_{q-1} \}_\alpha)
+\min \{ q-1,r-1 \}
\nonumber\\
&=&
q+2-r +\min \{q-1,r-1 \}
\\
&=&
\begin{cases}
q+1 & (r \leq q), \\
2q+1-r & (q+1 \leq r \leq q+2),
\end{cases}
\label{eq:DTPn}\end{eqnarray}
where in the last line we have taken account of eq.~(\ref{eqw:rleqq+2}).

For a single-component system ($q=1$), for example, this formula gives
$ 
D(T,P)=3-r,
$ 
which coincides with $f$ and is consistent with Fig.~\ref{fig:PD}.

\section{Conclusions}

Our principal results are
eqs.~(\ref{eq:Duvn}), (\ref{eq:DTvn}) and (\ref{eq:DTPn}).
We have also derived additional results, 
eqs.~(\ref{eq:Duvn_a}) and (\ref{eq:DTvn_a}).
For completeness, we have also described the known results, 
eqs.~(\ref{eq:Gibbs}) and (\ref{eq:Gibbs2}).
Here, 
eq.~(\ref{eq:Gibbs}) can also be written as
$ 
D(T,P,\mu_1, \ldots, \mu_q)=q+2-r,
$ 
because, by definition,
$f$ is the the dimension of a coexisting region in the space
that is spanned by the axes corresponding to $(T,P,\mu_1, \ldots, \mu_q)$ \cite{dependent}.

By collecting all these results, 
we obtain the following chain of equalities and inequalities;
\begin{eqnarray}
0
&\leq&
D(T,P,\mu_1, \ldots, \mu_q)
=q+2-r=f
\nonumber\\
&=&
D(T,P,\{ n^\alpha_1, \ldots, n^\alpha_{q-1} \}_\alpha)
=D(T,\{ v^\alpha, n^\alpha_1, \ldots, n^\alpha_{q-1} \}_\alpha)
\nonumber\\
&\leq&
D(T, P, n_1, \ldots, n_{q-1})
=q+1-r+\min \{q,r \}
\nonumber\\
&\leq&
D(T, v, n_1, \ldots, n_{q-1})
=q+2-r+\min \{q,r-1 \}
\nonumber\\
&\leq&
D(u, v, n_1, \ldots, n_{q-1})
=
D(\{ u^\alpha,v^\alpha,n^\alpha_1, \ldots, n^\alpha_{q-1}, \nu^\alpha \}_\alpha)=
q+1.
\label{eq:chain}\end{eqnarray}
Here, $q+1$ is 
the dimension of 
all possible states for a given value of $N_{\rm tot}$ \cite{Gibbs,AS,Callen}.
It also agrees with 
the dimension of the space
that 
is spanned by the axes corresponding to either 
$(T, P, n_1, \ldots, n_{q-1})$, 
or $(T, v, n_1, \ldots, n_{q-1})$,
or $(u, v, n_1, \ldots, n_{q-1})$.
This chain of equalities and inequalities 
may be regarded as the fundamental phase rule, 
which shows clearly how 
the dimension of a coexisting region varies depending on 
the choice of the variables.
It will be helpful in studying first-order phase transitions
and in drawing phase diagrams of new materials.

Finally, we note the following points.
Although we have assumed that 
the natural variables of entropy \cite{Gibbs,Landau,AS,Callen} 
are $U,V,N_1, \ldots, N_q$,
generalization to other cases 
(such as the case where they include the total magnetization
\cite{Callen,AS}) 
is straightforward.
Furthermore, we have assumed, as in the case of the Gibbs phase rule,
that there is no accidental degeneracy among 
equations 
which have been used in calculating $D$'s. 
Hence, it is in principle possible (though would be rare) 
that $D$'s take 
values that are different from our formulas.

\section*{Acknowledgment}

This work has been partly supported by KAKENHI (No. 19540415).


\begin{thebibliography}{99}
\bibitem{def:phase}
According to ref.~\citen{Fermi}, 
a phase is defined as a homogeneous part in 
a macroscopic system at equilibrium.
In contrast, 
some modern literature defines a phase 
according to the analytic properties of thermodynamical functions,
in imitation of the definition of `phase transition.' 
However, 
such a definition of phase 
would cause difficulties when, e.g., `coexistence of two phases' is 
discussed \cite{AS}.
About this point and precise definitions of `phase' and `phase transition,'
see Sec.~15.1 of ref.~\citen{AS}.
\bibitem{dependent}
These variables are not independent even in a region 
where only a single phase exists (i.e., $r=1$).
\end{thebibliography}
\end{document}